\documentclass[11pt]{article}
\usepackage{moriond,epsfig}

\def\Journal#1#2#3#4{{#1} {\bf #2}, #3 (#4)}

\def\PLB{{\em Phys. Lett.}  B}

\def\PRD{{\em Phys. Rev.} D}

\def\be{\begin{equation}}
\def\ee{\end{equation}}
\def\bea{\begin{eqnarray}}
\def\eea{\end{eqnarray}}

\begin{document}
\vspace*{4cm}
\title{SIMULATIONS OF
CLUSTERS OF GALAXIES}

\author{ S. SCHINDLER }

\address{Astrophysics Research Institute, Liverpool John Moores University,\\
Twelve Quays House, Birkenhead CH41 1LD, U.K.}

\maketitle\abstracts{
Numerical simulations of clusters of galaxies provide a unique way to
follow the 
dynamics of these systems. The models 
reveal many characteristics of the merging process of subclusters: 
shock structure and strength, temperature distribution and gas distribution.
From the models detailed observational signatures of the
dynamical state can be derived. 
The simulations also show that mergers have effects
on the magnetic field, on the X-ray luminosity, 
on the metal enrichment and other physical processes. 
Furthermore, observational methods
like the mass determination can be tested.
}

\section{Introduction}
Clusters of galaxies evolve on
time scales of the order or little less than the age of
the universe. Hence with observations one cannot follow their evolution.
Observations provide only snapshots of the different
evolutionary stages. Therefore only numerical simulations provide the
possibility to follow the dynamics of clusters.

Numerical models cannot only be used to learn how the
clusters form and evolve, but also observable quantities can be
derived, which can be compared directly with observations. This
comparison is used for many different purposes. Observations can be
interpreted more easily, e.g. the models can distinguish what observable
feature corresponds to which dynamical state.
Moreover, observational methods can be tested with the
simulated data. The results can be compared to the input parameters of
the calculations and in an iterative way the methods can be refined.
In this way the physical processes in clusters and cluster evolution
can be understood. This knowledge provides the opportunity to constrain
cosmological models.

Merging of subclusters are particularly 
interesting phenomena to
study with simulations. Irregular cluster morphologies in many X-ray
images of cluster as well 
as indications from optical observations imply that many clusters are
not relaxed. Hence mergers are very common in clusters of galaxies. 
Such mergers of subclusters are very energetic events, which affect
clusters strongly, e.g. the cause shocks. These shocks are the major
heating 
source for the intra-cluster gas and also particles can be
re-accelerated there to relativistic energies. 

This article is organised as follows. In Sect.~2 the different simulation
methods are explained. Sect.~3 shows what can be learned from  
models in particular about the merger process of clusters of galaxies. 
The effects 
of magnetic fields and other physical processes like radiative cooling
and star formation are discussed in Sections~4 and 5, respectively. In 
Sect.~6 results of simulations of different metal enrichment processes 
are presented. Sect.~7 shows how useful numerical models are to test 
the mass determination method. A short summary is given in Sect.~8.

\section{Simulation Methods}

For realistic simulations three-dimensional
calculations are required.  Also the different cluster
components must be taken into account: dark matter, 
galaxies and intra-cluster gas. Dark
matter and galaxies can be regarded as collisionless particles and
can therefore be modelled by N-body simulations. 
In these kinds of simulations only the gravitational interaction
between the particles are taken into account. Each particle is moved
in the force field of all the other particles. For current particle
numbers of $128^3$ or $256^3$ this procedure
would take a lot of computing time
to calculate the force by simply summing over all the other particles'
contributions. To accelerate the calculations different techniques
have been developed, e.g. the particles are 
sorted onto a grid or into a tree structure. In this way several 
particles are combined and treated simultaneously without loosing
much accuracy but gaining a lot of computing time. Many simulations
have been performed which take into account only dark matter and apply
therefore only N-body calculations. Such kind of simulations are very
useful for many purposes because the dark matter makes up most of the
gravitative mass. In this article, however, I will concentrate only on
models which include both, the dynamics of the dark matter and of the 
gas.

For the simulation of the gas pressure must be taken into
account, i.e. the full hydrodynamic equations must be solved. 
Two different
methods have been used generally for these hydrodynamic calculations: 
(1) Smoothed Particle Hydrodynamics (SPH; Lucy 1977; Monaghan 1985): 
This is a Lagrangian approach,
i.e. the calculation follows the fluid. The gas is treated as
particles in this approach. Examples of this type of simulations are:
Evrard (1990), Dolag et al. (1999), Takizawa
(1999), and Takizawa \& Naito (2000). 
(2) Grid-based codes: this is the Eulerian approach, i.e. the simulation volume is
divided into cells and  the fluid is
moving in this grid which is fixed in space. 
Examples of simulations
using grid codes can be found in Schindler \& M\"uller
(1993), Bryan et al. (1994), Roettiger et al. (1997),
Ricker (1998) and Quilis et al. (1998).

Fortunately, the choice of simulation technique is not essential. 
Calculations with both methods
yield very similar results. This was tested in a large project,
the Santa Barbara Cluster Comparison Project (Frenk et al. 1999), in
which the formation of a galaxy cluster was simulated using 12
different techniques. Both methods, SPH and grid codes, were applied.
Each simulation started with exactly the same initial conditions. The
comparison showed very good agreement in the properties of the dark
matter. Also relative good agreement was found in the gas temperature,
the gas mass fraction and the gas profiles of the final cluster. The
largest discrepancies were found in the X-ray luminosity which
differed by up to a factor of 2.  

\section{Cluster Models}

In general cluster models can reproduce cluster  morphologies and
parameters known
from observations quite well. The temperatures of the X-ray emitting
intra-cluster gas are typically very well simulated. Also the spatial
distributions are very realistic. For the dark matter component a NFW
profile (Navarro et al. 1995) is usally found, while the gas profile
is well fit by a so-called $\beta$-profile (Cavaliere \& Fusco-Femiano
1976). 

The models can  distinguish between cosmological parameters. 
Simulations on large scales show distinctly different distribution of
matter for a mean density $\Omega_M=1$ and $\Omega_M=0.3$,
respectively (Ostriker \& 
Cen 1996; Thomas et al. 1998; Jenkins et al. 1998). While in the latter
model the distribution changes only slightly between a redshift z=1
and now, in the $\Omega_M=1$ model significant differences are visible
in the same time interval: many smaller structures merge to larger 
structures, so that the distribution looks much less smooth at z=0 
than at z=1.
A distinction of $\Lambda=0$ models and
$\Lambda\ne0$ models is difficult though.

\subsection{Mergers of Clusters}

Simulations are ideal to follow the different stages of a merging event 
(see Fig.~1). In the 
pre-merger phase the two subclusters are approaching each other and
can still be
seen as well separated units. The collision, i.e. the moment when the cores
pass through each other, is characterized by enhanced X-ray luminosity
and enhanced temperature. In the subsequent post-merger
stage shock waves emerge -- mainly in direction of the original
collision axis. The shocks propagate outwards and are visible for
about 1 - 2 Gyrs.

\begin{figure}
\epsfig{figure=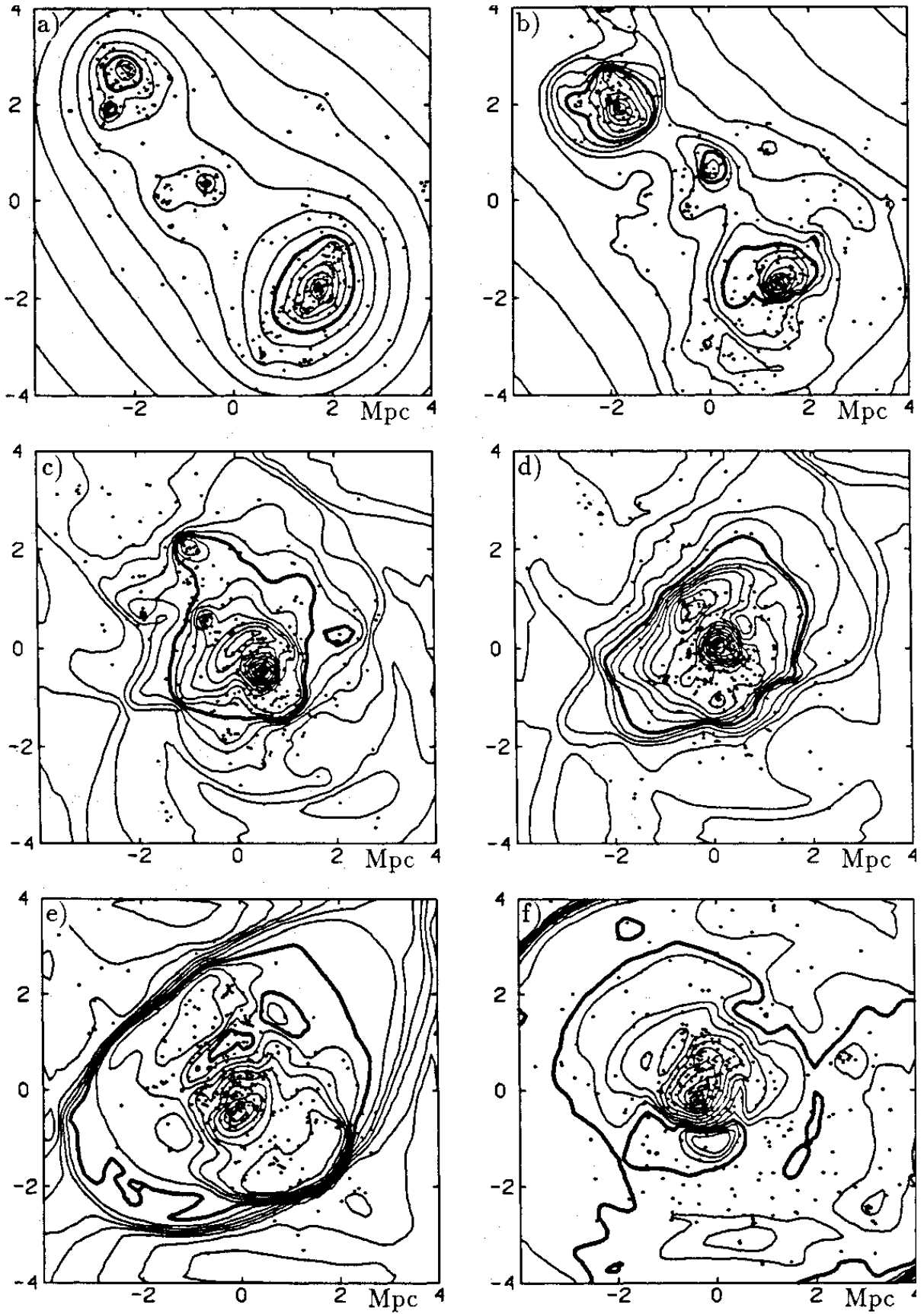,width=16cm,clip=}
\caption{Evolution of the X-ray temperature. The temperature contours
are logarithmically spaced with $\Delta \log T = 0.05$ 
the bold contour line corresponding to a temperature of 10$^8$ K. 
The six snapshots are taken at 
b) t = 0.95~Gyr, 
c) t = 2.7~Gyr,
d) t = 3.2~Gyr,
e) t = 4.4~Gyr,
f) t = 6.1~Gyr
after the configuration shown in a).}
\end{figure}

\subsection{Shocks}

The most prominent features emerging form mergers are shock waves in the 
intra-cluster gas resulting from colliding 
subclusters with relative velocities of up to $\approx$ 3000
km/s. When
a dense subcluster falls into a cluster a shock emerges
already before the core passage: a bow shock is 
visible in front of the infalling subcluster (Roettiger et al. 1997). 
The strongest shocks
emerge after the collision of subclusters, when these shocks propagate
(Schindler \& M\"uller
1993; Roettiger et al. 1999a; see Fig.~1e). These shocks are
relatively mild shocks, though, with a maximum Mach number of about 3. 
The shocks are visible as steep gradients in the gas density and in the gas
temperature. 
In general, the shock structure is found to be more filamentary at
early epochs of cluster formation
and quasi-spherical at low redshifts (Quilis et al. 1998).

Observationally, the shocks are best visible in X-ray temperature maps,
because they show up as steps in these maps (Fig.~1e; Schindler \& M\"uller
1993). For such maps
spatially resolved X-ray spectroscopy is necessary which can be
performed now with high accuracy with the new X-ray observatories XMM
and Chandra. 

The shocks are not only the major heat source of the intra-cluster
gas, but they are also of
particular interest for particle acceleration models. It is likely
that particles are re-accelerated to relativistic energies in these
shocks. These relativistic particles are responsible for 
the non-thermal
emission and their interaction with the cluster magnetic field shows
up as synchrotron emission in radio halos (e.g. Giovannini \&
Feretti 2001; Feretti 2001).

The shocks heat primarily the ions as has been shown in simulations which
treat ions and electrons separately (Chi\`eze et al. 1998; Takizawa 1999).
Only later the energy is transferred to the electrons.

\subsection{Observational Signatures}

Apart from shocks, there are several other signatures 
which can be used  to determine the dynamical
state of a galaxy cluster. For
example, the X-ray luminosity 
increases during the collision of two subclusters (Schindler \&
M\"uller 1993). The reason is that the gas is compressed, i.e. the gas
density is increased, and as the X-ray emission is proportional to the
square of the density we see enhanced X-ray emission during the
core passage of two subclusters. Shortly before the collision the gas
between the subclusters is heated due to compression and shows up as a high temperature region (see Fig.~1c). 

During the core passage and at each rebounce an increase in the magnetic
field is visible (Dolag et al. 1999; Roettiger et
al. 1999b; see also Section 4).
Mergers also cause turbulence and motion in the intra-cluster
gas. 
Off-centre collisions produce
additionally angular momentum (Ricker 1998; Roettiger et al. 1998). 

Observationally, mergers cannot only be identified by multiple X-ray
maxima, but also by isophote twisting with centroid shift and
elongations: the collisionless component is always elongated along the
collision axis, before and after the collision. 
The gas is first elongated along the collision
axis. During the core passage it is pushed out perpendicular to the
collision axis, so that later an elongation perpendicular to the
collision axis can be seen (Schindler \& M\"uller 1993; see Fig.~1e). 
Also offsets
between the collisionless component and the gas have been found
(Roettiger et al. 1997). 
If more than one merger occurs in a relatively short time interval the
temperature structure can become very complex.

\section{Cluster Simulations with Magnetic Fields}

Faraday rotation measurements indicate that clusters are permeated
by magnetic fields of the order of 1 $\mu$G  (e.g. Kim et al. 1991).
Also radio halos require the existence of magnetic fields in clusters 
on scales of a few
Mpc (e.g. Giovannini et al. 1991, 1993). 
Therefore magneto-hydrodynamic calculations have been performed
(Dolag et al. 1999; Roettiger et al. 1999b; see also Dolag, this
volume) 
to investigate the origin, the 
distribution and the 
evolution of the magnetic fields (see Fig.~2). Results of these
simulations are: 
the initial field distribution is irrelevant for the final structure of
the magnetic field. The structure is dominated only by the cluster
collapse. Faraday rotation measurements can be reproduced by the
simulations for
magnetic fields of the order of 1 $\mu$G in very good agreement with
the value inferred from observations. 

\begin{figure}
\epsfig{figure=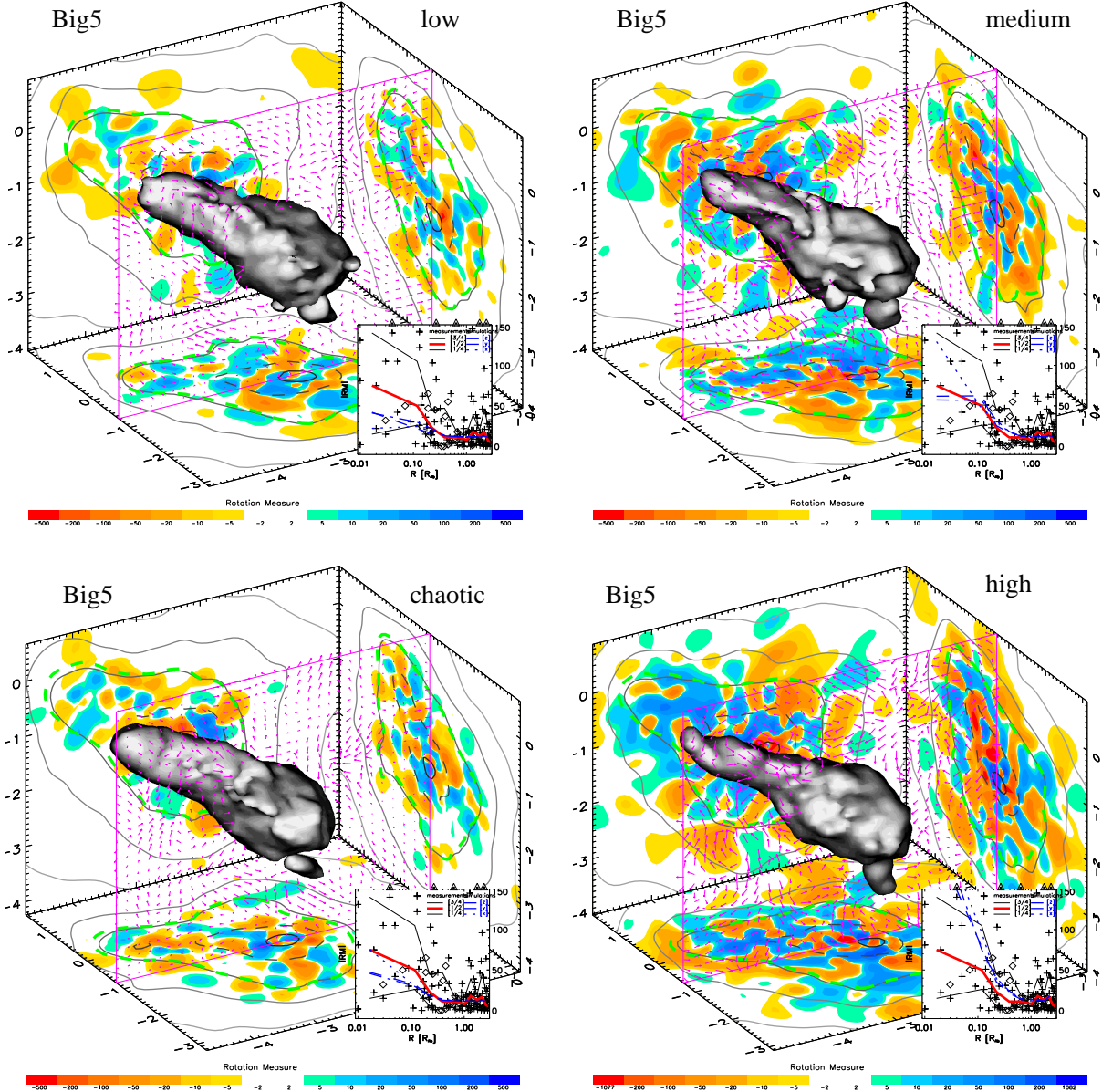,width=16cm,clip=}
\caption{Isodensity surface of four cluster models with 
magnetic fields in the core region of $0.4\mu G$ (``low'' and ``chaotic''), 
$1.1\mu G$ (``medium'') and $2.5\mu G$ (``high''), respectively,  
at redshift z=0. The
three projections show the rotation measure (colour) and the gas density 
(contours). The purple arrows indicate the
orientation and the strength of the magnetic field in a plane
containing the cluster centre. The insets show the rotation measure
versus distance from the cluster centre:  simulations (lines)
and observations in several clusters (symbols) 
(from Dolag (2000)).}
\end{figure}

Most important for the
amplification of the magnetic field are 
shear flows, while the compression of the
gas is of minor importance. Mergers  of subclusters
change the local magnetic field
strength as well as the structure of the cluster-wide field. At early
stages of the merger, filamentary structures prevail, which break
down later and leave a stochastically ordered magnetic field.

\section{Cooling and Star Formation}

It might be necessary to take into 
account 
additional physical processes 
in order to achieve realistic models. Two important 
processes are star formation and radiative cooling.

In simulations without cooling and star formation 
it was found, that the
gas is less concentrated than dark matter. Also the X-ray luminosity -
temperature relation inferred from simulations is 
in disagreement with observations (Eke et
al. 1998, Bryan \& Norman 1998, Yoshikawa et al. 2000).
The question arises, whether
this is a numerical artifact due to the negligence of physical processes 
or due to the difficulty in determining the X-ray luminosity correctly 
from the numerical models.
Another possibility is that 
it is connected with observational findings of different
profiles of baryonic and dark matter  (Schindler 1999) and the deviation
of the X-ray luminosity - temperature relation from a pure power law 
(Ponman et al. 1999) which could both be 
explained by non-gravitational heating processes. In order to test
this hypothesis
several groups have performed simulations with cooling and star 
formation and came to quite different conclusions.

Lewis et al. (2000) found that models with cooling and star formation  have a 
20\% higher X-ray luminosity and a  30\% higher
temperature in the cluster centre. 
Also Suginohara \& Ostriker (1998) find that 
radiative cooling increases luminosity.
In contrast to these results
Pearce et al. (2000) and Muanwong et al. (2001) find that radiative cooling 
decreases the total X-ray luminosity. 

In order to test whether the is gas less concentrated than the dark matter 
because of preheating, i.e. early non-gravitational heating, Bialek et
al. (2000) 
performed simulations with an initially elevated adiabat. They find that
they can reproduce the observations when adding 
an initial entropy of 55 - 150
keV cm$^2$.

Mathiesen \& Evrard (2001)  took a different approach  and tested with their 
models how good the observational temperature determination is. They 
simulated CHANDRA spectra and found that the temperature can be
underestimated by up to 20\% by the standard temperature determination 
method. The reason is cold material falling onto the cluster from all sides,
i.e. also along the line of sight, resulting in a cold contribution to the 
cluster spectrum and hence to a lower temperature determination.

Bryan \& Norman (1998) showed that in simulations the mass - temperature
relation is much more robust that the X-ray luminosity - temperature relation 
and hence suggested to use the former relation for drawing conclusions
about the cluster formation process.

The star formation rate can be affected by mergers of subclusters. It is 
still controversial whether mergers increase or decrease the star formation 
activity. The interstellar medium in 
a galaxy  can be compressed during a merger, which would
lead to an increased star formation rate. This effect was predicted 
by simulations by Evrard (1991). Also in a number of observations
a connection between mergers and enhanced star formation rate
has been found.
In contrast to these results
Fujita et al. (1999) found in simulations that the interstellar medium in the
galaxies is stripped off  due to increased ram pressure during 
the merger, so that the galaxies contain less gas and therefore show  
decreased star formation activity.  
At the moment of the subcluster collision,
they found an increase of 
post-starburst galaxies 
which indicates that a rapid drop in the star formation rate must have 
occured.

\section{Metal Enrichment} 

The intra-cluster  gas contains metals. 
This implies that the gas cannot be purely of primordial origin, but
it must at  
least partially have been processed in the cluster galaxies and  
have been 
expelled from the galaxy potential into the intra-cluster medium.
As mentioned above the star formation activity, i.e. the metal
production rate,  and its connection to the dynamical state 
is still controversial.
Also the gas ejection processes and their time scales are still under
discussion.  
Several gas ejection processes have been suggested: 
e.g. ram-pressure stripping 
(Gunn \& Gott 1972), galactic winds (De Young 1978), galaxy-galaxy
interaction and jets from active galaxies.

In order to decide which process is dominating at what time it is necessary to compare
observations and simulations in 
particular  with respect to the metallicity distribution within clusters and
to the metallicity evolution with redshift.

The effects of supernova driven
winds were studied by many groups. 
David  et al. (1991)
calculated the first models on cluster scales.
They found that the results depend sensitively on the input parameters:
the stellar initial mass function, the adopted supernova rate and 
the primordial fraction of intra-cluster gas. In 3D models calculating 
the gas dynamics and galactic winds
Metzler \&  Evrard (1994, 1997) found that
winds can account for the observed metal abundances. They find
very strong metallicity gradients (almost a factor  of ten between 
cluster centre and virial radius) which are hardly affected
by cluster mergers. From simulations on galaxy scale 
Murakami \& Babul  (1999) concluded that galactic winds are 
not very efficient for the metal enrichment. 

Another process which is probably important for the
metal enrichment is ram-pressure stripping: as a galaxy
approaches the cluster centre it experiences an increasing
pressure and at some point the galaxy potential is not
strong enough to retain the galaxy gas. The gas is stripped
off and the metals are released into the intra-cluster medium.

Simulations of ram-pressure are relatively difficult because 
not only the conditions of the gas inside the galaxy
and the potential of the galaxy must be taken into account, but
also the conditions of the surrounding medium. 
Recently, high resolution simulations were
carried out to study the stripping process in different types of
galaxies. Abadi et al. (1999) and 
Quilis et al. (2000) performed simulations of spiral galaxies. They
found that their gas can be stripped off  when it is
not homogeneous. Details of the spatial distribution and the time
dependence of the process are shown. For dwarf galaxies 
Mori \& Burkert (2000) found in their simulations that the gas is
easily stripped off  when these galaxies move through the intra-cluster
medium. In simulations of elliptical galaxies (Toniazzo \& Schindler 2001)
showed details of the stripping process of these galaxies. The amount of 
gas stripped off depends sensitively on the orbit of the galaxy.
The gas cannot only be stripped off as the galaxy approaches the
cluster centre, but the galaxy can again accumulate some gas when it is
in the apocentre of its orbit. X-ray morphologies of the
simulated galaxies are derived as well.

\begin{figure}
\centerline{\epsfig{figure=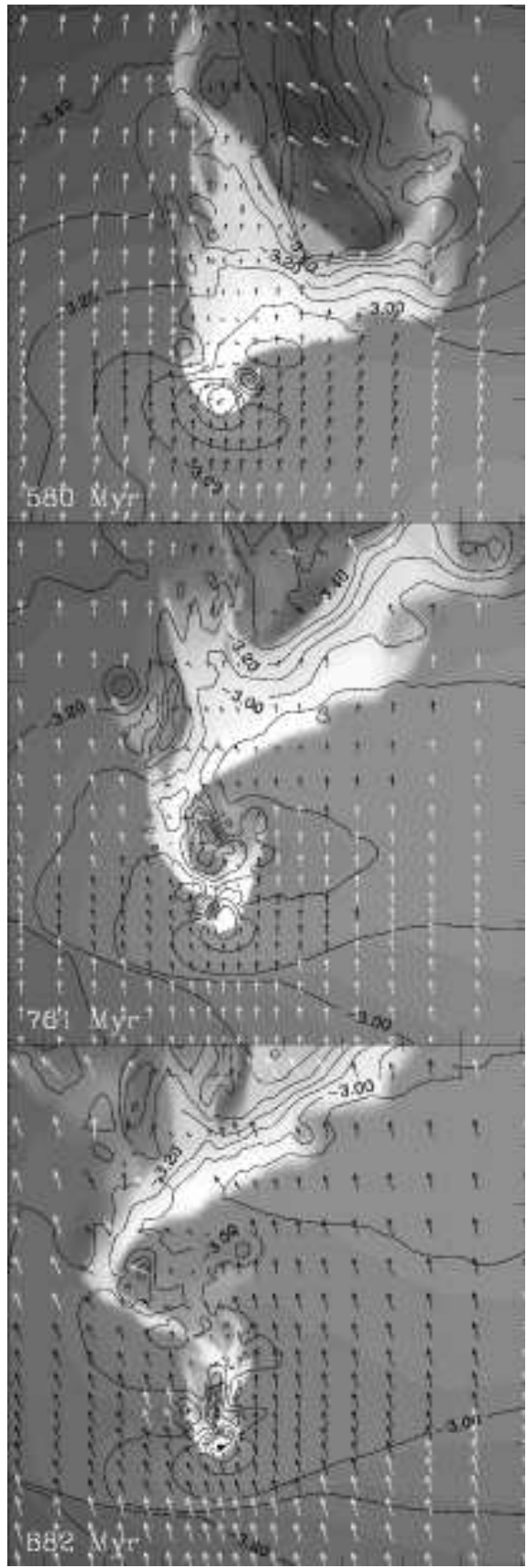,width=7.5cm,clip=}}
\caption{
Gas density (grey scale) and pressure (contours) of a galaxy
moving downwards towards the cluster centre. The arrows show the Mach
vectors (white when $M>1$, black otherwise). The gas of the galaxy is
stripped due to ram pressure (from Toniazzo \& Schindler (2001)). 
}
\end{figure}

All these simulations showed that ram-pressure stripping can be
an important metal enrichment process. Merging activity increases the
effect even more because the ram pressure is proportional to
$\rho_{ICM}\times v_{rel}^2$ with $v_{rel}$ being the relative
velocity of  intra-cluster gas and galaxies. During mergers not only the
gas density is increased but also the relative velocities are much higher
than in a relaxed cluster. Therefore a large influence of  the merging 
processes on the stripping rate are expected.

\section{Mass Determination of  Clusters of Galaxies}

Numerical  models are ideal tools to test observational methods. An example
for such a method is the mass determination from X-ray observations. 
In this method the X-ray emitting gas is used as a tracer for the potential.
Mass determinations are very important because they measure in an indirect way
the amount and the distribution of dark matter in clusters.

With the assumption of  hydrostatic equilibrium and spherical symmetry
the total cluster 
mass can be expressed in terms of 
only two observable quantities, which can both be inferred
from X-ray observations: the gas temperature and the gas density.

The total mass of a model cluster
is known exactly for each time step and within any radius.  On the other hand
the X-ray emission of the model cluster can be simulated and the normal
X-ray mass determination method can be applied to these simulated X-ray data.
A comparison of the true mass and the 
X-ray mass yields the accuracy of the X-ray method.

As shown by several groups the X-ray mass determination method proved to be quite
reliable in relaxed clusters (Evrard et al. 1996; Roettiger et al. 1996;
Schindler 1996). Typical errors are 15\% without any preference for
over- or underestimation.

Only during mergers quite strong deviations can occur.
The reason is that the two assumptions for the method -- 
hydrostatic equilibrium and spherical symmetry are not well  
obeyed during mergers. For example, at
the location of shocks the gas is not in hydrostatic
equilibrium. Shocks cause gradients -- both in the temperature and in
the density -- and can cause therefore an overestimation of the
mass. Locally, this can lead to a mass estimate up to two times the true
mass. Substructure on the other hand tends to flatten the azimuthally
averaged profile and hence leads to an underestimation of the mass, in
extreme cases to deviations of 50\% of the true mass (Schindler 1996).

In some cases, these deviations can be corrected for, e.g. in
clusters in which substructures are well distinguishable, the
disturbed part can be excluded from the mass analysis and a good
mass estimate can be obtained. But in general, mass determinations in
non-relaxed clusters should be done very cautiously.

Also the effect of magnetic fields on the mass determination was 
investigated. A considerable magnetic pressure compared 
to the thermal pressure would lead to
an underestimation of the mass.
Magneto-hydrodynamic simulations by Dolag et 
al. (1999) were used to perform the same comparison as mentioned above.
Dolag \& Schindler (2000) found that in relaxed clusters the mass
is underestimated at most by a few percent and only in the cluster
centre. In merging clusters, on the other hand,
the mass can be strongly underestimated. 
The reason is that during the merger the gas is compressed and with it the 
magnetic field lines and hence the magnetic field is stronger and affects the 
mass determination.

\begin{figure}
\epsfig{figure=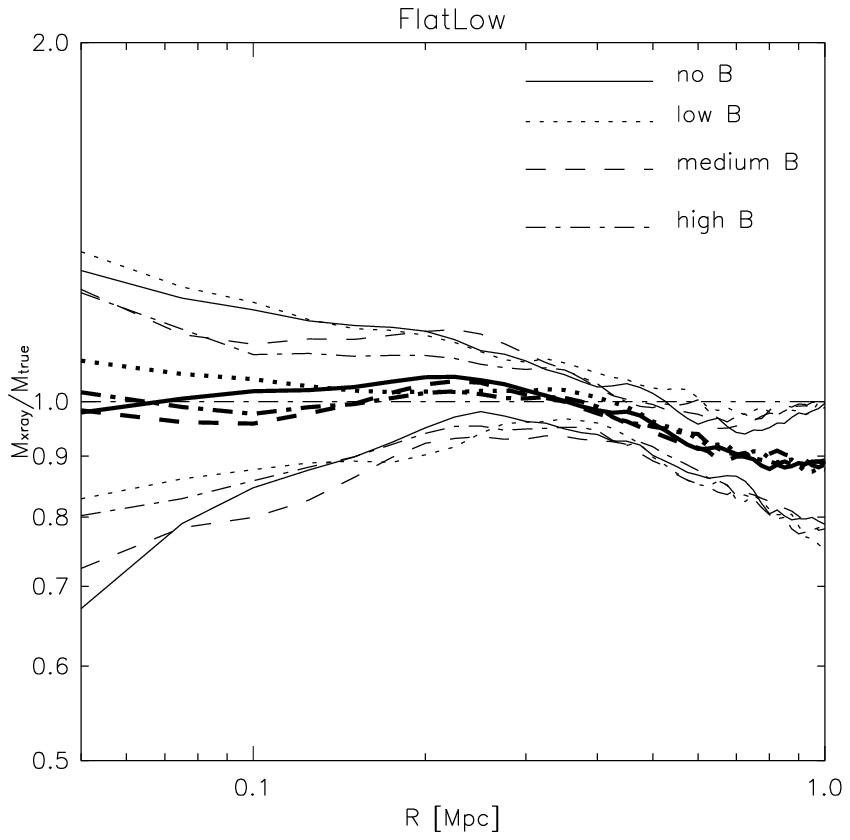,width=7cm,clip=}
\qquad
\epsfig{figure=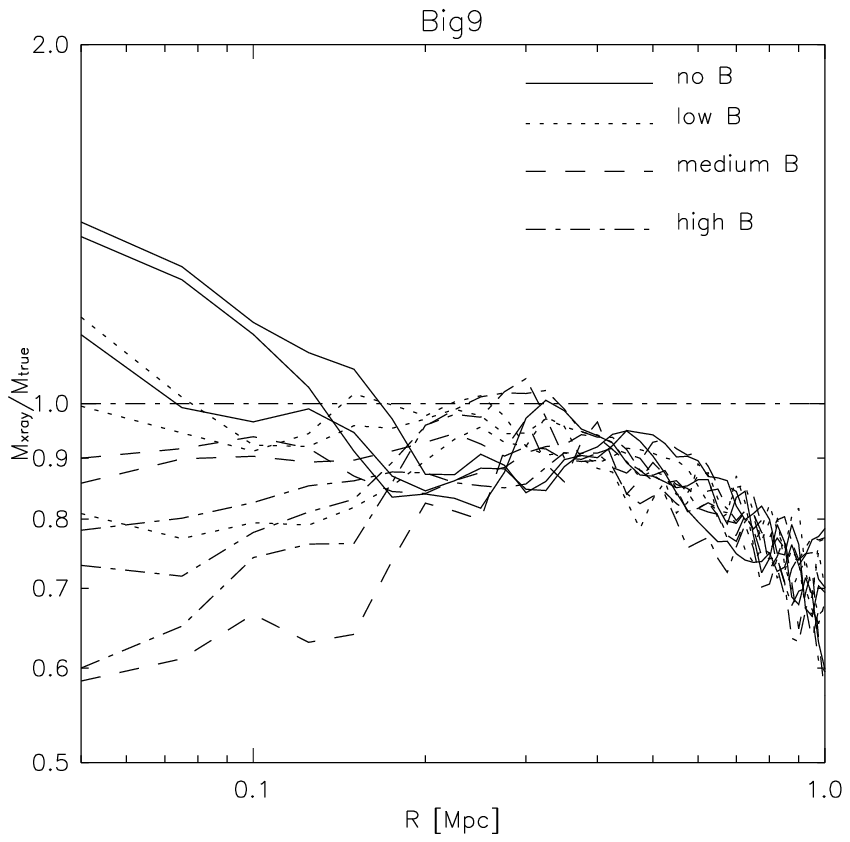,width=7cm,clip=}
\caption{Left:
Ratio of true and X-ray mass for 4 different
magnetic field strengths. The bold lines are averaged
profiles over 18 models, 
while the thin lines show the corresponding standard deviation.  
No dependence of the mass estimate on the magnetic field
strength is visible.
Right: Same for a cluster in the process of merging. For each
model, the three projection directions are shown in the same
line style. 
For this extreme merger model
the magnetic effects are considerably larger
than the usual scatter of mass profiles from different projection
directions (from Dolag \& Schindler (2000)).
}
\end{figure}

\section{Summary}

Simulations  provide a unique way to follow the dynamics of  galaxy clusters.
Different dynamical states can be distinguished.
Each stage of a merger (pre-merger, collision, post-merger) is characterised by
different observational characteristics.
Therefore not only merger clusters can be distinguished from
relaxed clusters but even the exact dynamical state can be determined by a 
detailed comparison of simulations and observations.

Numerical models can be used to test observational methods, like e.g. the
mass determination or the temperature determination. They define
the limitations of these methods and help improving them.

In the future not only the increase of resolution is important but also inclusion
of physical effects, e.g. radiative cooling, star formation,and magnetic 
field. All these improvements together will yield more realistic models, 
that will help us to understand the physical processes in clusters and to 
make the best use of clusters as diagnostic tools for cosmology. 

\section*{Acknowledgments}
I would like to thank the organisers for organising this extremely interesting, 
fruitful and enjoyable conference.


\section*{References}
\begin{thebibliography}{99}


\bibitem{} Abadi, M.G., Moore, B., Bower, R.G., 1999, MNRAS, 308, 947

\bibitem{} Bialek, J.J., Evrard, A.E., Mohr, J.J., 2000, astro-ph/0010584

\bibitem{} Bryan, G.L. Klypin, A., Loken, C.,
Norman, M.L., Burns, J.O., 1994, ApJ, 437, L5

\bibitem{} Bryan, G.L., Norman, M.L., 1998, ApJ, 495, 80

\bibitem{} Cavaliere A., Fusco-Femiano R., 1976, A\&A 49, 137

\bibitem{} Chi\`eze, J.-P., Alimi J.-M., Teyssier, R., 1998, ApJ, 495, 630

\bibitem{} David, L.P., Forman, W., Jones, C.,  1991, ApJ, 380, 39

\bibitem{} De Young, D.S., 1978, ApJ, 223, 47

\bibitem{} Dolag, K., 2000, Ph.D. Thesis,
Ludwigs-Maximilians-Universit\"at M\"unchen

\bibitem{} Dolag, K., Bartelmann, M., Lesch, H. 1999, A\&A, 348, 351

\bibitem{} Dolag, K., Schindler, S., 2000, A\&A, 364, 491

\bibitem{} Eke, V.R., Navarro, J.F., Frenk, C.S., 1998, ApJ, 503, 569

\bibitem{} Evrard, A.E. 1990, A\&A, 363, 349

\bibitem{} Evrard, A.E., 1991, MNRAS, 248, L8.

\bibitem{} Evrard, A.E., Metzler, C.A., Navarro, J.N. 1996, ApJ, 469, 494

\bibitem{} Feretti, L., Fusco-Femiano, R., Giovannini, G., Govoni, F., 2001, A\&A, in press, astro-ph/0104451

\bibitem{} Frenk, C.S., White, S.D.M., Bode, P., Bond, J.P., Bryan,
G.L., et al., 1999, ApJ, 525, 554

\bibitem{} Fujita, Y., Takizawa, M., Nagashima, M., Enoki, M.,
1999, PASJ, 51, L1

\bibitem{}  Giovannini, G., Feretti, L., Stanghellini, C., 1991,  A\&A,
252, 528

\bibitem{} Giovaninni, G., Feretti, L.,  Venturi, T., Kim, K.T., 
Kronberg, P.P., 1993, ApJ, 406, 399

\bibitem{} Giovaninni, G., Feretti, L., 2001, New Astronomy, in press,
astro-ph/0008342

\bibitem{} Gunn, J.E., Gott, J.R.III, 1972, ApJ, 176, 1

\bibitem{} Jenkins, A., Frenk, C.S., Pearce, F.R., Thomas, P.A.,
Colberg, J.M., et al., 1998, ApJ 499, 20

\bibitem{} Kim, K.T., Tribble, P.C., Kronberg, P.P., 1991, ApJ,
379, 80

\bibitem{} Lewis G.F., Babul, A., Katz, N., Quinn, T., Hernquist, L.,
Weinberg, D.H., 2000, ApJ, 536, 623

\bibitem{} Lucy, L.,  1977, AJ, 82, 1013

\bibitem{} Mathiesen, B.F., Evrard, A.E., 2001, ApJ, 546, 100

\bibitem{} Metzler, C.A., Evrard, A.E., 1994, ApJ, 437, 564

\bibitem{} Metzler, C.A., Evrard, A.E., 1997, astro-ph/9710324

\bibitem{} Monaghan, J.J., 1985, Comp. Phys. Rept., 3, 71 

\bibitem{} Mori, M., Burkert, A., 2000, ApJ, 538, 559

\bibitem{} Muanwong, O., Thomas, P.A., Kay, S.T., Pearce, F.R.,
Couchman, H.M.P., 2001, astro-ph/0102048

\bibitem{} Murakami, I., Babul, A., 1999, MNRAS, 309, 161

\bibitem{} Navarro, J.F., Frenk, C.S., White, S.D.M., 1995, MNRAS 275, 720

\bibitem{} Ostriker, J., Cen, R., 1996, ApJ, 464, 27

\bibitem{} Pearce, F.R., Thomas, P.A., Couchman, H.M.P., Edge, A.C.,
2000, MNRAS, 317, 1029

\bibitem{} Ponman, T.J., Cannon, D.B., Navarro, J.F., 1999, Nature 397, 135

\bibitem{} Quilis, V., Ib\'a\~nez, J.M., S\'aez, D., 1998, ApJ, 502, 518

\bibitem{} Quilis, V., Moore, B., Bower, R., 2000, Science, 288, 1617

\bibitem{} Ricker, P.M., 1998, ApJ, 496, 670

\bibitem{} Roettiger, K., Burns, J.O.,  Loken, C., 1996, ApJ, 473, 651

\bibitem{} Roettiger, K., Burns, J.O., Stone J.M., 1999a, ApJ, 518, 603

\bibitem{} Roettiger, K., Loken, C., Burns, J.O., 1997, ApJS, 109, 307

\bibitem{} Roettiger, K., Stone J.M., Burns, J.O., 1999b, ApJ, 518, 594

\bibitem{} Roettiger, K., Stone J.M., Mushotzky, R.F., 1998, ApJ, 493, 62

\bibitem{} Schindler, S., 1996, A\&A, 305, 756

\bibitem{} Schindler, S., 1999, A\&A 349, 435

\bibitem{} Schindler, S., M\"uller, E., 1993, A\&A, 272, 137

\bibitem{} Suginohara, T., Ostriker, J.P., 1998, ApJ, 507, 16

\bibitem{} Takizawa, M., 1999, ApJ, 529, 514

\bibitem{} Takizawa, M., Naito, T., 2000, ApJ, 535, 586

\bibitem{} Toniazzo, T., Schindler, S., 2001, MNRAS, in press, astro-ph/0102204

\bibitem{} Thomas, P.A., Colberg, J.M., Couchman, H.M.P., Efstathiou,
G., Frenk, C.S., 1998, MNRAS, 296, 1061

\bibitem{} Yoshikawa, K., Jing, Y.P., Suto, Y., 2000, ApJ, 535, 593

\end{thebibliography}
\end{document}